\documentclass[prb,aps,twocolumn,groupedaddress,superscriptaddress,eqsecnum]{revtex4-2}
\usepackage{ulem}
\usepackage[usenames,dvipsnames]{color}
\usepackage[colorlinks=true,linkcolor=Blue,citecolor=Blue,urlcolor=Blue]{hyperref}
\usepackage{graphicx}
\usepackage{verbatim}
\usepackage{bbm}
\usepackage{bm}
\usepackage{amsmath} 
\usepackage{amssymb}
\usepackage{mathptmx}
\usepackage[version=3]{mhchem}
\usepackage{float}

\newcommand{\ba}{\begin{eqnarray}}
\newcommand{\ea}{\end{eqnarray}}
\newcommand{\bd}{\begin{displaymath}}

\newcommand{\nn}{\nonumber \\}

\newcommand{\cpn}{${\rm CP}^{N-1}$~}

\begin{document}
\title{Dynamics of $\mathrm{CP}^{N-1}$ skyrmions}
\author{Seungho \surname{Lee}}
\email{seungho6626@gmail.com} 
\affiliation{Department of Physics, Sungkyunkwan University, Suwon 16419, South Korea}
\affiliation{Institute of Basic Science, Sungkyunkwan University, Suwon 16419, South Korea}
\author{Hyojae \surname{Jeon}}
\email{wjsgywo980@gmail.com}
\affiliation{Department of Physics, Sungkyunkwan University, Suwon 16419, South Korea}
\author{Jung Hoon \surname{Han}}
\email{hanjemme@gmail.com}
\affiliation{Department of Physics, Sungkyunkwan University, Suwon 16419, South Korea}
\begin{abstract} We derive several exact results for the dynamics of CP$^{N-1}$ skyrmions with arbitrary $N$. Fractonic continuity equation is shown to hold for arbitrary CP$^{N-1}$ fluid implying the conservation of the topological charge and the dipole moment. Inclusion of the Gilbert damping modifies the continuity equation, resulting in the violation of the dipole moment conservation but not of the topological charge. Thiele's equation for the CP$^{N-1}$ skyrmion follows from the modified continuity equation. The Girvin-MacDonald-Platzman (GMP) algebra in the long-wavelength limit is derived for arbitrary CP$^{N-1}$ fluid. In the case of CP$^2$ skyrmions, we construct a finite-size skyrmion with dominant ferromagnetic moments. We find that its magnetic dipole texture carries a nonzero CP$^1$ charge equal to half the CP$^2$ skyrmion charge and argue that the topological Hall effect of electrons should exist due to their coupling to the ferromagnetic part of the CP$^2$ texture.
\end{abstract}
\date{\today}
\maketitle

\section{Introduction}
Conventional treatments of magnetism and magnetic dynamics are framed in terms of O(3) classical spins or SU(2) quantum spins. Extensions of the scheme to include orbital and multipolar degrees of freedom in a spin-orbit-coupled system also exist. More recently, interesting developments have taken place to understand magnetic dynamics in multipolar systems~\cite{batista21,dahlbom22,batista22} with quadrupolar and higher-order multipolar moments~\cite{ivanov_2008_prl,shannon_2013_prb,shannon22,zhang_2024_prl,shannon_2024_prb}. In a concurrent effort, extension of the theory of CP$^1$ skyrmions to CP$^{N-1}$ ($N>2$) has taken place~\cite{akagi_jhep_2021,akagi21,amari22,batista23NC,amari24,batista25,hayashi2025} with emphasis on the {\it structure} of \cpn skyrmions in multipolar magnets~\cite{akagi21,amari22,batista23NC,amari24}. In this work, we explore various aspects of the {\it dynamics} of the \cpn fluid and of the isolated \cpn skyrmions. Along the way, several known results for CP$^1$ fluids such as the fractonic conservation laws~\cite{papanicolaou91,garst24} and topological Hall effect~\cite{Nagaosa13,han17} of skyrmions are extended to \cpn. 

Magnetic skyrmions have been discovered in a host of ferromagnets~\cite{pfleiderer09,Yu10,Yu2018,skyrmionics20}. The skyrmions discovered to date are the CP$^1$ skyrmions defined by the homotopy $\pi_2 ({\rm CP}^1) = \mathbb{Z}$. In spin-1 magnets, on the other hand, the five magnetic quadrupolar moments as well as the three magnetic dipole moments together form an eight-component order parameter, represented by CP$^2$ field. The possibility of CP$^2$ skyrmion formation in spin-1 magnets has been scrutinized recently~\cite{akagi21,amari22,batista23NC}. A suggestion to create CP$^3$ skyrmions in a magnetic bilayer also exists~\cite{batista25}. It is becoming clear that \cpn theory should provide a unifying framework to address magnetic dynamics of multipolar magnets. 

A strong analogy among the dynamics of electrons in the lowest Landau level, vortices in superfluids, and skyrmions in magnets has been noted for a long time~\cite{thouless93,han17}. In a modern perspective, they can be viewed as manifestations of fractonic dynamics and hydrodynamics~\cite{gromov21,garst24} in diverse physical platforms. The Girvin-MacDonald-Platzman (GMP) algebra, originally conceived as a manifestation of the constrained dynamics of electrons in the lowest Landau level~\cite{GMP}, has found close connection to fractonic field theory~\cite{son22} and the theory of CP$^1$ fluid~\cite{garst24}. 

In this paper, we show that the fractonic continuity equation for the CP$^1$ fluid, originally derived  in the early 90s~\cite{papanicolaou91}, generalizes to CP$^{N-1}$ with arbitrary $N>2$. The violation of the fractonic conservation and the concomitant loss of dipole conservation, but not the topological charge conservation, in the presence of Gilbert damping is derived. The GMP algebra obeyed by the CP$^1$ topological density~\cite{garst24} is shown to generalize to arbitrary $N$. These findings are reported in Sec.~\ref{sec:EoM&GMP}. The question of the topological Hall effect arising from mobile electrons coupled to \cpn skyrmion textures is addressed in Sec.~\ref{sec:STT}, with particular emphasis on CP$^2$ skyrmions. Within the BPS variational framework, we construct finite-size ferromagnetic CP$^2$ skyrmions, whose characteristic magnetic structure is further supported by numerical energy minimization beyond the BPS ansatz. These skyrmions are expected to give rise to the topological Hall effect. Some mathematical backgrounds are summarized in Sec.~\ref{sec:background} and in several appendices.

\section{mathematical backgrounds}
\label{sec:background}

{ The fundamental field that we denote simply as the CP$^{N-1}$ field $\bf z$ arises from the projective map
\begin{align}
    [\mathbf z]:\mathbb{R}^{2+1}\longrightarrow \mathrm{CP}^{N-1},
\end{align}
from the (2+1)-dimensional spacetime $x= ({\bf r}, t)$ to the complex projective space
\begin{align}
     \mathrm{CP}^{N-1}\simeq \mathrm{SU}(N)/(\mathrm{SU}(N-1)\times \mathrm{U}(1)).
\end{align}
Specifically, $[\mathbf z]$ denotes the equivalence class of nonzero vectors $\mathbf z\in\mathbb C^N$ under the local identification $\mathbf z(x)\sim e^{i\chi(x)}\mathbf z(x)$ for arbitrary function $\chi$. The fields are normalized to $\mathbf z^\dagger\mathbf z=1$. 

The Hamiltonian $H$ under consideration is local, translationally invariant, and  invariant under the local $\mathrm{U}(1)$ gauge transformation ${\bf z} \rightarrow e^{i\chi} {\bf z}$. In other words,  
\begin{align}
    H[ {\bf z} ] =
    \int d^2\mathbf r\, \mathcal H\left(\mathbf z,\mathbf z^\dagger,
    D_i\mathbf z,(D_i\mathbf z)^\dagger\right),
\end{align}
where the energy density $\mathcal H$ is a function of ${\bf z} ~ ({\bf z}^\dag)$ and covariant derivatives ($\mu = t, x,y$) 
\begin{align} D_\mu  {\bf z} &= P_{\mathbf z}\partial_\mu  {\bf z} = (\partial_\mu - ia_\mu) \mathbf{z} , 
\end{align}
with the gauge field 
\begin{align} a_\mu = -i\mathbf{z}^\dagger\partial_\mu \mathbf{z} 
\end{align}
and the projector
\begin{align} 
P_{\mathbf z} = \mathbf{1}_{N\times N}- {\bf z} {\bf z}^\dag  . 
\end{align} 

The Hamiltonian $H[{\bf z}]$ can be written equivalently using the real-valued order parameter ${\bf n}$ of fixed length, related to ${\bf z}$ by
\begin{align}
    n^a = {\bf z}^\dag \lambda_a {\bf z} 
\end{align}
for $a=1 \cdots N^2 -1$, and $\lambda_a$ are the generators of SU($N$) obeying the algebra~\cite{Kaplan1967}
\begin{align}\label{product}
    \lambda_a \lambda_b = \frac{2}{N} \delta_{ab} \mathbf{1}_{N\times N} + (d_{abc} + i f_{abc})\lambda_c ,
\end{align}
with the structure constants $f_{abc}$, the symmetric coefficients $d_{abc}$, and the normalization following from
\begin{align}\label{ids-lambda}
    f_{abc} &= \frac{1}{4 i} \mathrm{Tr}\left(\lambda_a [ \lambda_b, \lambda_c]\right)\,, \nn
    d_{abc} &= \frac{1}{4} \mathrm{Tr}\left(\lambda_a \{ \lambda_b, \lambda_c \}\right)\,,
    \nn
    \delta_{ab} &= \frac{1}{2}\mathrm{Tr}\left(\lambda_a \lambda_b\right)\,.
\end{align}
The order parameter $n^a$ obey the constraints
\begin{align}
 n^a n^a =  \frac{2(N-1)}{N}\,, ~~~ 
    n^a = \frac{N}{2(N-2)} d_{abc} n^b n^c\,. 
    \label{n-constraint}
\end{align}
The proof of the identities~\eqref{n-constraint} is given in Appendix~\ref{proof-n}.
Hereafter, repeated indices are implicitly summed over. The use of ${\bf n}$ instead of $\bf z$ facilitates the connection of our theory to familiar language in magnetism expressed in terms of magnetic dipoles, quadrupoles, and so on. Decomposition of ${\bf n}$ into various magnetic multipoles will be elaborated later. 

The topological three-current for \cpn can be defined as
\begin{align}
    J^\mu = \frac{1}{2\pi} \epsilon^{\mu\nu\lambda} \partial_\nu a_\lambda = \frac{1}{8\pi} \epsilon^{\mu\nu\lambda}f_{abc} n^a \partial_\nu n^b \partial_\lambda n^c  
    \label{Jmu-in-n-and-z} 
\end{align}
in both CP$^{N-1}$ and real order parameters. The CP$^{N-1}$ definition in particular highlights the topological nature of the conservation law $\partial_\mu J^\mu =0$. The current $J^\mu$ is given by the Hodge dual of the Fubini-Study K\"ahler two-from, and the conservation law automatically follows from the closedness of the two-form.

Throughout the paper we adopt the metric signature $(+,-,-)$ in relating $J^\mu$ to $J_\mu$. Another useful identity we use is 
\begin{align}  \partial_\mu n^a \partial^\mu n^a  = 4 ( D_\mu {\bf z} )^\dag ( D^\mu {\bf z} ) 
\label{Identity}
\end{align}
to relate the magnetic energy density in terms of ${\bf n}$ and ${\bf z}$. The integral of the temporal component is the CP$^{N-1}$ topological charge 
\begin{align}
\label{top_charge}
Q_{N-1} = \int_\mathbf{r} ~ J_t \quad \left(\int_\mathbf{r} := \int d^2 \mathbf{r} \right)\,,
\end{align} 
which is an integer by virtue of the second homotopy $\pi_2 ({\rm CP}^{N-1}) = \mathbb{Z}$. 

The equation of motion for the complex field $\mathbf{z}$ is given by the Hamiltonian flow
\begin{align}
    iD_t\mathbf z=P_{\mathbf z}\frac{\delta H}{\delta\mathbf z^\dagger} . 
    \label{z-eom-main}
\end{align}
This equation is invariant under the local $\mathrm{U}(1)$ transformation $\mathbf z\mapsto e^{i\chi}\mathbf z$. The same equation of motion can be expressed in terms of ${\bf n}$ as the generalized Landau-Lifshitz (gLL) equation~\cite{batista21,dahlbom22,batista22}
\begin{align}
    \label{gLL}
    \partial_t {n}^a= 2 f_{abc}\frac{\delta H}{\delta n^b} n^c  .
\end{align}
Both \eqref{z-eom-main} and \eqref{gLL} follow from the action
\begin{align}
    S & = i \int d^2 {\bf r} dt ~ {\bf z}^\dag {\partial_t} {\bf z} - \int dt H \nn 
    & = \frac{1}{2} \int d^2 {\bf r} dt  \int_0^1 du f_{abc}n^a{\partial_t n^b \partial_u n^c} - \int dt H \, . 
    \label{gWZR}
\end{align}
The derivation of the gLL equation from the action $S$ is given in App.~\ref{app:gLLG}. The familiar LL equation for the O(3) magnet is recovered by substituting  $f_{abc} =\varepsilon_{abc}$.}

\section{Fractonic conservation law and GMP algebra}
\label{sec:EoM&GMP}

\subsection{Conservation laws} 
In the absence of damping, the three-current {$J^\mu$} given in \eqref{Jmu-in-n-and-z} in terms of either ${\bf z}$ or ${\bf n}$ satisfies the continuity equation {$\partial_\mu J^\mu =0$} for any \cpn magnet - see Appendix~\ref{app:topological-current-cons} for explicit derivation. As emphasized by Skyrme~\cite{skyrme61,skyrme62}, conservation laws that exist independent of the symmetry of the action is deemed {\it topological}. We can go further and re-write continuity equation for \cpn fluid as the {\it fractonic} { continuity equation} ($i,j=x,y$)
\begin{align} \partial_t J_t + \partial_i \partial_j J_{ij} = 0 \, ,
\label{fractonic-conservation} \end{align}
where a tensor current
\begin{align} J_{ij} = {-\frac{\epsilon_{ik}}{2\pi} \sigma_{kj}} \end{align} 
is defined in terms of the stress tensor {
\begin{align} \sigma_{kj} &= \mathcal{H}\delta_{kj} -  \frac{\partial\mathcal{H}}{\partial(\partial_j n^a)} \partial_k n^a \nonumber
\\
&= \mathcal H\delta_{kj}
        -2\,\mathrm{Re}\left(
        \left(\frac{\partial\mathcal H}{\partial(D_j\mathbf z)^\dagger}\right)^\dagger
        D_k\mathbf z
        \right).
\label{stress-tensor} 
\end{align} 
Derivations are given in App.~\ref{app:topological-current-cons} and App.~\ref{app:z-fractonic} for the expression in terms of ${\bf n}$ and $\mathbf z$, respectively.}

The fractonic continuity equation \eqref{fractonic-conservation} implies the conservation of the total dipole moment ${\bf D}$ in addition to the total charge:
\begin{align} {\bf D} = \int_{\bf r} ~ {\bf r} J_t . \end{align}
Due to the dipole conservation, a single \cpn skyrmion cannot move; instead only the skyrmion-antiskyrmion pair (a skyrmion dipole) with a fixed relative position can move freely. Two skyrmions of the same charge can rotate around each other while preserving their center of mass. The fractonic conservation law in \eqref{fractonic-conservation} for $N=2$ had been known since the early 90s~\cite{papanicolaou91} though the word `fracton' did not exist at the time. A similar fractonic conservation law exists for the vortices in superfluids as well~\cite{gromov21}, highlighting the generality of the fractonic description for these states of matter. We have here shown that all \cpn  fluids obey the fractonic dynamics { {\it regardless of} the specific nature or the symmetries (except that of translation) of the governing energy functional $H$.} 

Including the Gilbert damping modifies the dynamics given by \eqref{gLL} and, accordingly, the fractonic conservation law. The gLL equation in the presence of Gilbert damping becomes

\begin{align}\label{gLLG}
\partial_t n^a & = 2f_{abc}\frac{\delta H}{\delta n^b}n^c - \alpha \frac{\delta H}{\delta n^a}  \nn 
& = 2f_{abc}\frac{\delta H}{\delta n^b}n^c {+ \alpha f_{abc} f_{cde} n^b n^d \frac{\delta H}{\delta n^e}} , 
\end{align}
where $\alpha >0$ represents damping~\cite{gilbert,batista22}. The continuity equation \eqref{fractonic-conservation} is modified to
\begin{align}\label{modified-cont-eq}
    \partial_t J_t + \partial_i \partial_j J_{ij} & = \frac{\alpha}{4\pi} \epsilon_{ij} f_{abc} \partial_i\left[\frac{\delta H}{\delta n^a} n^b \partial_j n^c\right] .
    \nn
    &  = \frac{\alpha}{8\pi} \epsilon_{ij}\partial_i\left[\partial_t n^c \partial_j n^c\right] + O(\alpha^2) . 
\end{align}
For derivation of the modified continuity equation, see Appendix~\ref{app:topological-current-cons}. The second line follows from the first by invoking the gLL equation \eqref{gLL}. Under the new continuity equation \eqref{modified-cont-eq}, the topological charge $Q_{N-1}$~\eqref{top_charge} is still conserved but the dipole moment is not:
\begin{align} \frac{d}{dt} {\bf D} 
&= -\frac{\alpha}{8\pi} \hat{x} \int_\mathbf{r} \partial_t \mathbf{n} \cdot \partial_y \mathbf{n}  + \frac{\alpha}{8\pi} \hat{y} \int_\mathbf{r} \partial_t \mathbf{n} \cdot \partial_x \mathbf{n} . 
\label{dDdt}
\end{align} 

Further progress can be made for the {\it rigid} configuration, ${\bf n}({\bf r} , t) = {\bf n}({\bf r} - {\bf R}(t))$, such that $\partial_t {\bf n} = - ( \dot{\bf R} \cdot \nabla ) {\bf n}$. Inserting this back into \eqref{dDdt} and assuming $\int_{\bf r} ( \partial_x {\bf n} )^2 = \int_{\bf r} (\partial_y {\bf n} )^2 = I$ and $\int_{\bf r} \partial_x {\bf n} \cdot \partial_y {\bf n} = 0$ as in a rotationally symmetric skyrmion gives an equation of motion for the dipole moment
\begin{align}
\dot{\bf D} = \frac{\alpha  I}{8\pi} \dot{\bf R} \times \hat{z} \implies \hat{z}\times \dot{\bf D} = \frac{\alpha  I}{8\pi} \dot{\bf R} . 
\end{align}
For a rigid skyrmion we can also write ${\bf D} = Q_{N-1} {\bf R}$, $\dot{\bf D} = Q_{N-1} \dot{\bf R}$, and recover Thiele's equation~\cite{thiele, oleg2022}
\begin{align} Q_{N-1}\hat{z}\times \dot{\bf R} = \frac{\alpha  I}{8\pi} \dot{\bf R} + {\bf F}_{\rm ext}
\end{align}
in the absence of the external force ${\bf F}_{\rm ext}$. For a single \cpn skyrmion, even allowing for dissipation gives $\dot{\bf R}=0$ unless there is an external force ${\bf F}_{\rm ext}$ acting on it. This well-known result for CP$^1$ skyrmion is now shown to hold for arbitrary \cpn. 

A more dramatic implication of the modified conservation law \eqref{modified-cont-eq} is in regard to the creation or annihilation of the \cpn skyrmion-antiskyrmion pair, which changes the dipole moment ${\bf D}$ and would not be possible under the strict fractonic conservation law of \eqref{fractonic-conservation}. The creation of skyrmion-antiskyrmion pair, or a skyrmion dipole, had been frequently seen in numerical simulations of LLG equation~\cite{Tchoe_Han_2012, Fert2013}, but its connection to the non-conservation of the dipole moment through the equation \eqref{dDdt} had not been noted before. 

\subsection{GMP algebra}
A convenient way to capture the dynamics of the \cpn order parameters $n^a$ is through the Poisson bracket $\partial_t n^a = \{ n^a , H \}$ along with the fundamental relation
\begin{align}
    \{ n^a ({\bf r} ), n^b ({\bf r}'  ) \} = 2f_{abc} n^c ({\bf r} ) \delta ({\bf r} - {\bf r}' )\,, 
    \label{Poisson-bracket-for-n} 
\end{align}
which is the classical correspondence to the commutation relation of the $\mathrm{SU}(N)$ generators $[\lambda_a, \lambda_b] = 2 i f_{abc} \lambda_c $~\cite{batista21}. With a given energy functional $H$, the gLL equation \eqref{gLL} follows from the Poisson bracket as 
\begin{align} \frac{\partial n^a}{\partial t} = \{n^a  , \, H \} = \int_{\bf r} \frac{\delta H}{ \delta n^b } \{n^a,n^b\}= 2 f_{abc} \frac{\delta H }{ \delta n^b}  n^c. \end{align} 
%
One can show, based on the fundamental relation \eqref{Poisson-bracket-for-n}, that 
\begin{align} \{ J_t ({\bf r} ) , {\bf n} ({\bf r}' ) \} =  {-\frac{\epsilon_{ij}}{2\pi}\partial_i {\bf n} \partial_j \delta ({\bf r}-{\bf r}' )}  \end{align}
and
\begin{align} 
\{J_t({\bf r} ),J_t({\bf r}' )\} = {-\frac{\epsilon_{ij}}{2\pi} \partial_i J_t({\bf r} ) \partial_j \delta( \bf r - \bf r' )}\,
\label{cpn-GMP}
\end{align}
for the topological charge density $J_t$---see Appendix~\ref{app:GMP} for derivation. This is the Girvin-MacDonald-Platzman (GMP) algebra, first derived for the density operators in the lowest Landau level to describe neutral excitations~\cite{GMP}, for the topological density operators in the \cpn fluid. A similar algebra had been constructed for CP$^1$ fluid~\cite{garst24}, vortices in superfluids~\cite{volovik79}, and for the fractonic field theory~\cite{son22}. The Poisson bracket for the dipole moment follows from the GMP algebra \eqref{cpn-GMP} as
\begin{align}
    \{D_i, D_j\} = { \frac{\epsilon_{ij}}{2\pi} Q_{N-1}}\,.
\end{align}

To summarize, we have derived the fractonic conservation law for the topological three-current and its modification under damping, as well as the GMP algebra for the topological density, valid for arbitrary \cpn fluid.



\section{Topological Hall effect of CP$^2$ skyrmion} 
\label{sec:STT}

The theory of spin transfer torque is based on the strong Hund coupling of local spin texture ${\bf S}$ to the itinerant spin moment $\psi^\dag {\bm \sigma} \psi$, both of which transform as O(3) vectors and can be coupled linearly. The two-component spinor $\psi = (\psi_\uparrow ~ \psi_\downarrow )^T$ represents electrons of both spin orientations. 

In \cpn magnets, the $(N^2 -1)$-component vector $n^a$ breaks down as three magnetic, five quadrupolar, seven octupolar components and so on totalling $N^2 -1$ components. To couple higher-order local moments to mobile electrons, one must first form products of electron spins such as $(\psi^\dag \sigma_a \psi) (\psi^\dag \sigma_b \psi) \cdots$ with the indices $a,b = x,y,z$ arranged to form a matching multipole moment. Such interactions are inevitably of higher order in the Hund's interaction strength and are likely to grow weaker with the increasing order of multipoles. A reasonable assumption is that even in multipolar magnets, the main interaction channel between electron fluids and the local moments is through their dipolar Hund coupling. It is thus essential to inquire if a \cpn skyrmion texture with $N>2$ can also carry nonzero CP$^1$ topological charge. We address this question by constructing a quite general CP$^2$ skyrmion texture and examining its CP$^1$ charge. 

A well-known ``trick" for constructing a model CP$^1$ configuration is to use the Bogomol’nyi-Prasad-Sommerfield ({ BPS}) inequality~\cite{Bogomolny1976,Prasad1975,Belavin1975,belavin1975metastable,Lichnerowicz1970}. We first show that this trick applies for arbitrary \cpn and use this result to construct CP$^2$ skyrmion configurations of arbitrary charge~\cite{hayashi2025}. The inequality reads 
\begin{align} 
    (D_{x} {\bf z} + i {\rm sgn}(Q_{N-1}) D_{y}{\bf z})^\dag (D_{x} {\bf z} + i {\rm sgn}(Q_{N-1}) D_{y}{\bf z}) \geq 0\label{equality1},
\end{align}  
for \cpn fields ${\bf z}$. The sign of the \cpn charge $Q_{N-1}$ is explicitly included. Equivalently,
\begin{align}
     \sum_{j=x,y} |D_j {\bf z} |^2 \geq  \frac{{\rm sgn}(Q_{N-1})}{i}   \left((D_x {\bf z})^\dagger D_y{\bf z} 
    -(D_y {\bf z})^\dagger D_x{\bf z}\right) . 
\end{align}
Invoking the identity \eqref{Identity} and \eqref{Jmu-in-n-and-z} and integrating over all space, 
\begin{align}
E \equiv \frac{1}{2}\int_\mathbf{r} \sum_{j =x,y} (\partial_j {\bf n})^2 \geq 4\pi | Q_{N-1} | .
\end{align}
For a given charge $Q_{N-1}$, the energy $E$ is minimized by achieving the equality $E = 4\pi | Q_{N-1}| $ when 
\begin{align} D_x {\bf z} + i{\rm sgn}(Q_{N-1}) D_y {\bf z} = 0. 
\end{align} 
This equation is solved by
\begin{align}
    {\bf z} = \frac{\xi^{|n|}{\bf u} + r^{|n|}e^{in\theta}{\bf v}}{(\xi^{2|n|} + r^{2|n|})^{1/2}}\label{general_solution},
\end{align}
where $n$ is the integer-valued \cpn skyrmion charge $Q_{N-1}$, $\xi$ is the skyrmion radius, and ${\bf u}$ and ${\bf v}$ are two arbitrary $N$-dimensional complex unit vectors satisfying ${\bf u}^* \cdot {\bf v} = 0$~\cite{ivanov_2008_prl,DADDA1978,Longinov23}. One can work out the corresponding gauge field 
\begin{align}
    {\bf a} = \frac{n r^{2(|n|-1)}}{\xi^{2|n|}+r^{2|n|}}\left( - y , x \right) 
\end{align}
and the covariant derivatives $D_\mu {\bf z} = (\partial_\mu - ia_\mu ) {\bf z}$ to show that $D_x {\bf z} + i {\rm sgn} (n) D_y {\bf z} = 0$, and that the topological number is indeed $n$:
\begin{align}
    Q_{N-1} = \frac{2\xi^{2|n|}n|n|}{2\pi} \int \frac{r^{2|n|-1}}{(\xi^{2|n|}+r^{2|n|})^2}drd\theta = n . 
\end{align}
The \cpn charge is $n$ regardless of the choice of ${\bf u}, {\bf v}$ vectors, { or the skyrmion radius $\xi$}. 

Specializing to CP$^2$, the skyrmion should have textures in both dipolar and quadrupolar sectors. The question we address now is, given the CP$^2$ charge $n$, would the texture also support nonzero $Q_1$? One must first keep in mind that, unlike the CP$^2$ charge, the CP$^1$ charge of the skyrmion solution \eqref{general_solution} depends on the choice of ${\bf u}, {\bf v}$. Rather than making arbitrary choices for the two unit vectors, we use energy consideration to narrow down the scope of $({\bf u}, {\bf v})$ vectors. For the energy functional, the previous energy
\[
E=\frac12\int \sum_j(\partial_j{\bf n})^2
\]
is supplemented with an anisotropy {and an SU($N$)-invariant quartic derivative term}:
\begin{align}\label{energy_with_quadratic}
    E'=E+\kappa \int_{\bf r}{\bf S}^2{+\eta\int_{\bf r}\left[(D_i{\bf z})^\dagger D_i{\bf z}\right]^2
    }.
\end{align}
where the magnetic dipole ${\bf S}=(S^x,S^y,S^z)$ can be worked out from ${\bf z}$ in \eqref{general_solution} by
${\bf S}={\bf z}^\dag(\lambda_7,-\lambda_5,\lambda_2){\bf z}$
(see Appendix~\ref{app:CP2-algebra} for details of CP$^2$ algebra).
{
Our approach is to employ the BPS solution given in Eq.~\eqref{general_solution} as a variational ansatz for the extended energy functional. Within this ansatz, the anisotropy term lifts the degeneracy associated with the internal vectors $({\bf u},{\bf v})$, and the competition between the anisotropy and the SU($N$)-invariant quartic derivative term stabilizes the skyrmion size. Since the quartic derivative term affects only the skyrmion size within the BPS ansatz (see Appendix~\ref{app:energy-analysis}), the determination of the optimal pair $({\bf u},{\bf v})$ is governed solely by the anisotropy term.
}
The task at hand is thus to find a pair of $({\bf u}, {\bf v})$ that minimizes the functional
\begin{align}\label{minimize the energy}
\kappa \!\int_\mathbf{r}\!\frac{\xi^{4|n|}4|\mathbf{u}_r\!\times\!\mathbf{u}_i|^2+r^{4|n|}4|\mathbf{v}_r\!\times\!\mathbf{v}_i|^2+2\xi^{2|n|}r^{2|n|}\left(1-|\mathbf{u}\!\cdot\!\mathbf{v}|^2\right)}{(\xi^{2|n|}+r^{2|n|})^2},
\end{align}
where $\mathbf{u}_r$, $\mathbf{u}_i$, $\mathbf{v}_r$, and $\mathbf{v}_i$ are real vectors defined by $\mathbf{u}=\mathbf{u}_r+i\mathbf{u}_i$ and $\mathbf{v}=\mathbf{v}_r+i\mathbf{v}_i$.

{
For $\kappa>0$, minimizing the anisotropy energy formally requires $({\bf u},{\bf v})=(e^{i\alpha/2},e^{i\beta/2})\hat{\bf e}$ for arbitrary phases $(\alpha,\beta)$ and a real-valued unit vector
$\hat{\bf e}$ (see Appendix~\ref{app:energy-analysis} for details of the energy minimization). Such a choice yields
${\bf S}=0$, corresponding to a purely quadrupolar texture. However, this configuration is incompatible with the BPS constraint ${\bf u}^* \cdot {\bf v}=0$ imposed by Eq.~\eqref{general_solution}. We therefore conclude that no purely quadrupolar CP$^2$ skyrmion exists within the BPS ansatz when $\kappa >0$.}

\begin{figure}[t]
\normalsize
  \includegraphics[width=\columnwidth]{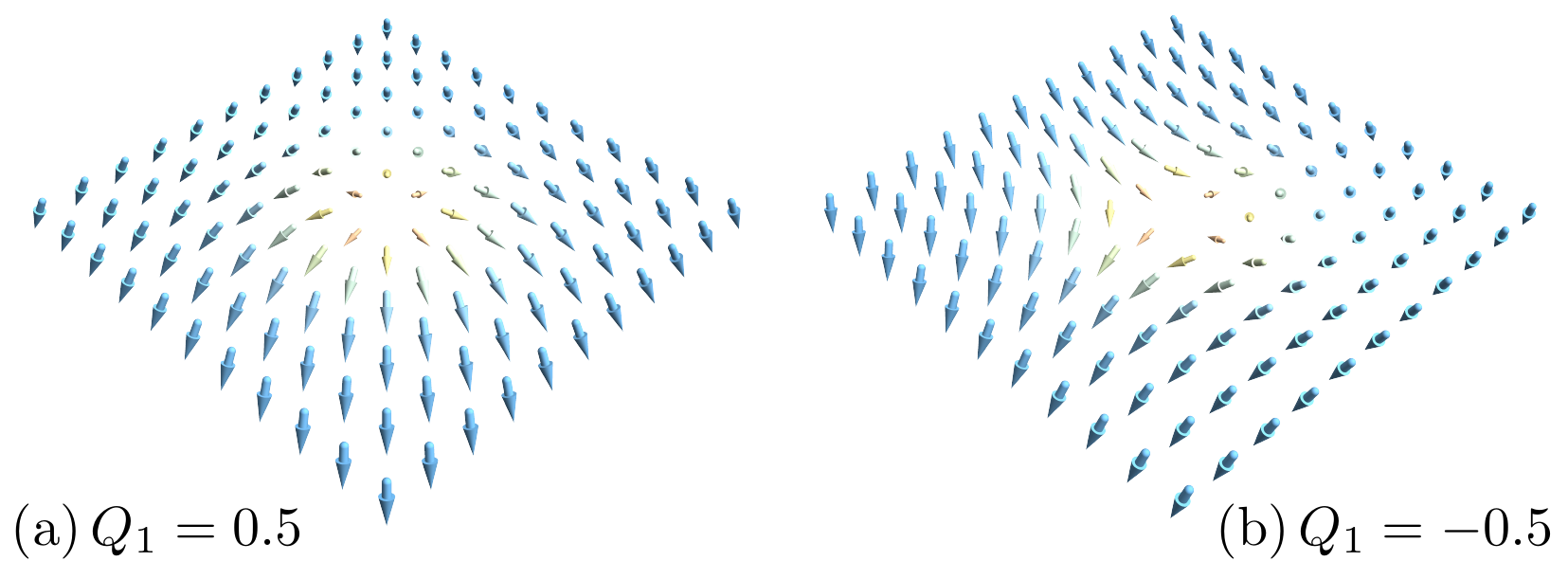}
  \caption{Magnetic dipole texture ${\bf S}$ for $\alpha=\pi/4$, $\beta=\pi$, $\gamma=\pi/4$, $\varphi=0$, and $\xi=1$ in \eqref{general_simple_function}. (a) $n=1$, (b) $n=-1$.}
  \label{figure}
\end{figure}

For $\kappa < 0$, as shown in Appendix~\ref{app:energy-analysis}, {the minimizer of the ansatz} is parameterized by
\begin{align}
    {\bf u} \!=\! e^{i\varphi}
    \begin{pmatrix}
        \cos{\alpha}\sin{\beta}\\
        \sin{\alpha}\sin{\beta}\\
        \cos{\beta}
    \end{pmatrix},
    \qquad
    {\bf v} \!=\! \frac{e^{-i\gamma}}{\sqrt{2}}
    \begin{pmatrix}
        \cos{\alpha}\cos{\beta}-i\sin{\alpha}\\
        \sin{\alpha}\cos{\beta}+i\cos{\alpha}\\
        -\sin{\beta}
    \end{pmatrix}
\label{general_simple_function}
\end{align}
with arbitrary angles $(\alpha,\beta,\gamma,\varphi)$. The magnetic dipole and quadrupole profiles associated with such $({\bf u},{\bf v})$ can be worked out (see Appendix~\ref{app:CP2-algebra}), and the CP$^1$ skyrmion number following from its magnetic dipole texture is
\begin{align}
Q_1 &= \frac{1}{4\pi}\int_{\bf r}
\frac{{\bf S}\cdot\partial_x{\bf S}\times\partial_y{\bf S}}
{|{\bf S}|^3}\nn\\
&=\frac{1}{4\pi}\int
\frac{2n|n|\xi^{2|n|}r^{|n|-1}}
{(r^{2|n|}+2\xi^{2|n|})^{3/2}}
drd\theta
=\frac{n}{2},
\label{Q1-for-CP2-skyrmion}
\end{align}
which is half the CP$^2$ skyrmion number.

The reason for the half-integer value of $Q_1$ above can be seen in Fig.~\ref{figure}, which shows the magnetic dipole configuration ${\bf S}$ for $n=\pm1$. At first sight it is tempting to view these half-integer entities as merons when $n$ is an odd integer, but a closer examination reveals a more complicated picture. The size of the magnetic dipole moment is given by
\begin{align}
{\bf S}^2=
\frac{r^{2|n|}(r^{2|n|}+2\xi^{2|n|})}
{(r^{2|n|}+\xi^{2|n|})^2}.
\end{align}
independent of the angles parameterizing $({\bf u},{\bf v})$. The size of the magnetic dipole moment is depleted as $r^{|n|}$ near the origin and reaches unity for $r/\xi\gtrsim1$, { where the skyrmion radius $\xi$ 
\begin{equation}\xi
=
\left[\frac{2\eta|n|^2(|n|+1)}{3|\kappa|}\right]^{1/4}.
\end{equation}
This skyrmion radius is fixed by the competition between the quartic derivative term and the anisotropy term  introduced in \eqref{energy_with_quadratic} for the BPS ansatz \eqref{general_solution}. For derivation of the formula, see App. \ref{app:energy-analysis}.}

Thus for $\kappa<0$, the CP$^2$ skyrmion texture becomes mostly magnetic outside the core and predominantly quadrupolar inside. We refer to such texture as the {\it ferromagnetic} CP$^2$ skyrmion, to emphasize its largely magnetic character outside the core. A typical meron, on the other hand, would have the spin pointing along $\pm\hat z$ and winding around the equator at infinity, costing infinite magnetic energy. The origin of the half-integer value of $Q_1$ in Eq.~\eqref{Q1-for-CP2-skyrmion} is attributed to the vanishing of the magnetic dipole moment at the origin, which is forbidden in the CP$^1$ skyrmion but is allowed for CP$^2$ skyrmion.

\begin{figure}[t]
    \centering
    \includegraphics[width=\columnwidth]{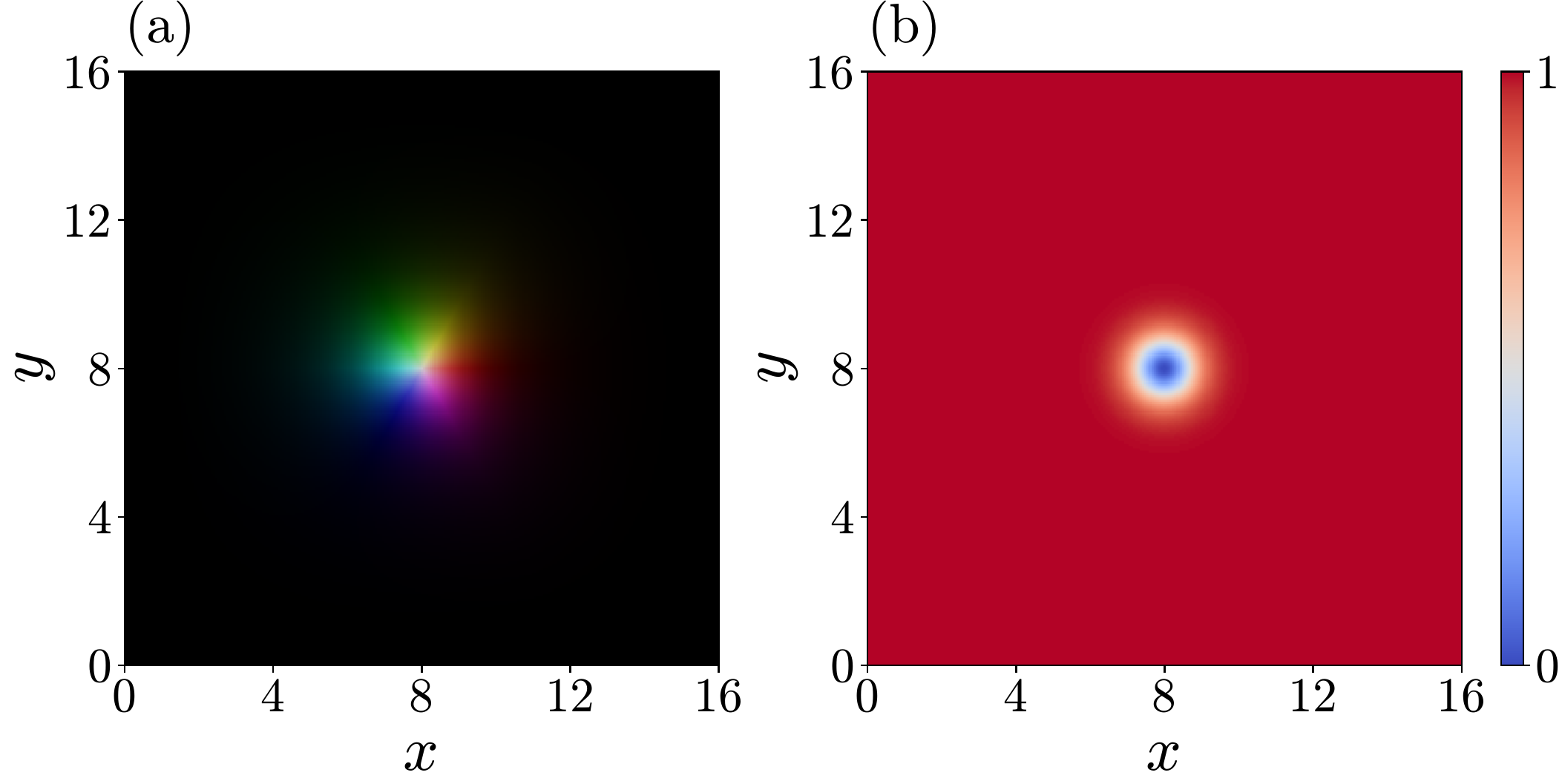}
    \caption{%
    \normalsize
    \leftskip=0pt
    \rightskip=0pt
    \parfillskip=0pt plus 1fil
    Ferromagnetic CP$^2$ skyrmion obtained by minimizing
    the energy functional in Eq.~\eqref{energy_with_quadratic}.
    (a) Magnetic dipole texture with colors representing
    the local spin direction. Black indicates the south pole
    $\mathbf{S}=(0,0,-1)$ and hue represents the azimuthal
    angle of the dipole. The bright gray at the core indicates $\mathbf{S} = (0,0,0)$.
    (b) Spatial distribution of the squared magnetic dipole
    moment $|\mathbf S|^2$, which vanishes at the skyrmion core.
    }
    \label{fig:numerical}
\end{figure}

To examine whether the characteristic properties of the ferromagnetic CP$^2$ skyrmion persist beyond the BPS variational ansatz,
we perform numerical energy minimization of Eq.~\eqref{energy_with_quadratic} in the full CP$^2$ configuration space via the arrested Newton flow~\cite{Speight2020,Gudnason2020,Gudnason2022,Leask2022,Leask2026}. The relaxed configuration retains a CP$^2$ skyrmion charge $Q_2\approx 1$ and a magnetic dipole skyrmion charge
$Q_1\approx 1/2$, in agreement with the analytical results. As shown in Fig.~\ref{fig:numerical}, the magnetic dipole moment vanishes at the skyrmion core.

The Hund coupling between local and itinerant moments and the theory of topological Hall effect derived from such coupling assumes that the magnitude $|{\bf S}|$ remains finite, which will be the case for the ferromagnetic CP$^2$ skyrmion outside the core. We thus expect that ferromagnetic CP$^2$ skyrmions carrying nonzero $Q_1$ exhibit the topological Hall effect despite the appearance of magnetic quadrupolar moments in the core.

\section{Summary and discussion}

Our central findings concerning the fractonic continuity equation, its modification by the Gilbert damping, and the GMP algebra obeyed by the CP$^{N-1}$ topological density remain valid independently of the detailed form of the translation-invariant Hamiltonian.

We also investigate $\mathrm{CP}^{2}$ skyrmions using a variational approach and full-field numerical calculations. Our results support the existence of an isolated $\mathrm{CP}^{2}$ skyrmion with a magnetic dipole texture carrying a $\mathrm{CP}^{1}$charge of $1/2$.

\acknowledgments
JHH was supported by the National Research Foundation of Korea (NRF) grant funded by the Korea government (MSIT) (Grant No. 2023R1A2C1002644 and No. RS-2024-00410027). We acknowledge delightful conversations with Dung Nguyen Xuan on the field-theoretic understanding of GMP algebra and its relation to fractonic hydrodynamics. SL thanks Seunghun Lee for discussions. { JHH thanks Aleksander Głódkowski and Jie Wang for insightful comments on the paper.} 

{ Note added: After the submission of this work, we became aware of the work by Brauner {\it et al}~\cite{Brauner} which emphasized the general validity of the fractonic conservation law for dynamical systems governed by symplectic form of the action with translation invariance. Many physical systems obey such conditions, ferromagnets, superfluids. The $\mathrm{CP}^{N-1}$ fluids now constitute an additional physical platform for which these general arguments are applicable. The influence of damping and the consequent destruction of dipole conservation is new.}

\appendix

\section{Constraints on the order parameter}\label{proof-n}
{Here we prove the algebraic identity \eqref{n-constraint} satisfied by the order parameter $\bf n$. Let us introduce a projector $l\equiv \mathbf{z}\mathbf{z}^\dagger$, i.e., $l_{ab} = z_a z^*_b$. The normalization $\mathbf{z}^\dagger \mathbf{z} = 1 $ directly gives
\begin{align}
    \mathrm{Tr} (l)= 1\,, \quad l^2 = l\,.
\end{align}
Since \(l=\mathbf{z}\mathbf{z}^\dagger\) is Hermitian, we expand it as
\begin{align}
    l=\alpha \mathbf{1}_{N\times N}+\beta_a\lambda_a .
\end{align}
From \(\mathrm{Tr}(l)=1\) and \(\mathrm{Tr}(\lambda_a)=0\), we obtain \(\alpha=1/N\).
Multiplying by \(\lambda_b\) and taking the trace, Eq.~\eqref{ids-lambda} gives
\begin{align}
    \mathrm{Tr}(\lambda_b l)=2\beta_b.
\end{align}
On the other hand, using \(l=\mathbf{z}\mathbf{z}^\dagger\),
\begin{align}
    \mathrm{Tr}(\lambda_b l)=\mathbf{z}^\dagger\lambda_b\mathbf{z}=n^b.
\end{align}
Hence \(\beta_b=n^b/2\), and therefore
\begin{align}
    l=\frac{1}{N}\mathbf{1}_{N\times N}+\frac12 n^a\lambda_a.
\end{align}
Using Eq.~\eqref{product}, we obtain
\begin{align}
    l^2
    = \left(\frac{1}{N^2} + \frac{1}{2N} n^a n^a \right)\mathbf{1}_{N\times N}
    + \left(\frac{1}{4} d_{abc} n^a n^b + \frac{1}{N} n^c \right)\lambda_c .
\end{align}
Imposing \(l^2=l\) and comparing the coefficients of \(\mathbf{1}_{N\times N}\) and \(\lambda_c\), we obtain
\begin{align}
    \frac{1}{N^2} + \frac{1}{2N} n^a n^a &= \frac{1}{N}, \nn
    \frac{1}{4} d_{abc} n^a n^b + \frac{1}{N} n^c &= \frac{1}{2} n^c .
\end{align}
This is equivalent to Eq.~\eqref{n-constraint}.
}

\section{Generalized Landau-Lifshitz-Gilbert equation}
\label{app:gLLG}

The dynamics of $\mathrm{CP}^{N-1}$ skyrmions stems from the gLL, which is derived from the Poisson bracket with the Hamiltonian $\partial_t n^a = \{n^a, \, H\} = \int \delta H/ \delta n^b \{n^a,n^b\}= 2f_{abc}(\delta H/ \delta n^b) n^c$.
In this section, we show the gLL can be formally introduced from the $\mathrm{CP}^{N-1}$ Wess-Zumino action which is defined as
\begin{align}
    S_\mathrm{WZ}=\int_{x^3} \int_0^1 du \frac{1}{2}f_{abc}n^a{\partial_t n^b \partial_u n^c}\,.
\end{align}
{where $\int_{x^3}= \int d^2 {\bf r} dt$, and $n^a( x^\mu, u)$ is a smooth extension of the original field $n^a( x^\mu)$ such that $n^a( x^\mu, 0) = n^a_0$ (a constant) and $n^a( x^\mu, 1) = n^a( x^\mu)$. The variation of the Wess-Zumino action is given by
\begin{align}
  \delta S_\mathrm{WZ}
  & = \frac{1}{2}\int_{x^3} \int_0^1 du f_{abc} \big[{3\delta n^a \partial_t n^b \partial_u n^c}  \nonumber
  \\
  & \quad \quad \quad \quad {+ \partial_t \left( n^a \delta n^b \partial_u n^c  \right) + \partial_u \left( n^a \partial_t n^b \delta n^c \right)} \big] \nonumber
  \\
  & = {\frac{1}{2}\int_{x^3} \int_0^1 du f_{abc}  \partial_u \left( n^a \partial_t n^b \delta n^c  \right)} \nonumber
  \\
  & = {-\frac{1}{2}\int_{x^3}  f_{abc}  n^a \delta n^b \partial_t n^c}  \,.
\end{align}
Note that here we assume the condition $\delta n^a (x^\mu, 0) = 0,\, \delta n^a(x^\mu, 1) = \delta n^a (x^\mu)$ and thereby the integration of the total derivative with respect to $u$ does not vanish. The corresponding functional derivative from the variation of $S_\mathrm{WZ}$ is
}
\begin{align}
    \frac{\delta S_\mathrm{WZ}}{\delta n^a} = {-\frac{1}{2}f_{abc} \partial_t n^b n^c}\,.
\end{align}
The equation of motion follows from the extremization of the total action $S = S_\mathrm{WZ} - \int dt H$: 
\begin{align}
     {-f_{abc} \partial_t n^b n^c} &= 2\frac{\delta H}{\delta n^a}\,.
     \label{gLL-v1}
\end{align}
One can multiply the left-hand side of Eq.~\eqref{gLL-v1} by $f_{ade} n^d$ and invoke some identities  
\begin{align}\label{id}
\partial_\mu n^a & = \frac{N}{(N-2)} d_{abc} n^b \partial_\mu n^c \nn 
f_{abc}f_{ade}  & = {\frac{2}{N}\left(\delta_{bd}\delta_{ce} - \delta_{be}\delta_{cd}\right) + d_{bda}d_{cea} - d_{bea}d_{cda}}
\end{align}
to get the simplified expression ${f_{abc}f_{ade} \partial_t n^b n^c n^d = -\partial_t n^e}$. Multiplying the same factor to the right-hand side of Eq.~\eqref{gLL-v1} gives the gLL in the conventional form:
\begin{align}
    \partial_t n^e &= 2 f_{ade} \frac{\delta H}{\delta n^a} n^d\,.
\end{align}

The dynamics of \cpn skyrmions in the presnece of dissipation can be captured by the generalized Landau-Lifshitz-Gilbert (gLLG) equation
\begin{align}
    \partial_t {n}^a &= 2 f_{abc}\frac{\delta H}{\delta n^b} n^c - \alpha \frac{\delta H}{\delta n^a} \nonumber
    \\
    &= 2 f_{abc}\frac{\delta H}{\delta n^b} n^c {+ \alpha f_{abc} f_{cde} n^b n^d \frac{\delta H}{\delta n^e}}
\end{align}
The equivalence of the damping terms in the first and second lines can be proved by employing the identity
\begin{align}\label{id_functional_derivative}
    \frac{\delta H}{\delta n^a} = \frac{N}{N-2} d_{abc} n^b \frac{\delta H}{\delta n^c}\,.
\end{align}
Equation~\eqref{id_functional_derivative} stems from the constraint [Eq.~\eqref{n-constraint}]. From the constraint of the field, we directly obtain the constraint for tangent vectors
\begin{align}\label{tangent-constraint}
    \dot{n}^a = \frac{N}{N-2} d_{abc} n^b \dot{n}^c\,,
\end{align}
where $\dot{n}$ is a derivative with respect to an arbitrary parameter. Any tangent vector can be expressed as a linear combination of elements of a basis of the tangent space
\begin{align}
    \dot{n}^a &=  C^\alpha e_\alpha^a
\end{align}
for some coefficients $C^\alpha$ and basis $\{\mathbf{e}_1, \mathbf{e}_2, \cdots, \mathbf{e}_{N^2-1}\}$.
Then, from Eq.~\eqref{tangent-constraint}, we have
\begin{align}\label{id_tangent}
     C^\alpha e_\alpha^a &= \frac{N}{N-2} d_{abc} n^b  C^\alpha e_\alpha^c \nonumber
    \\
    e_\alpha^a &= \frac{N}{N-2} d_{abc} n^b e_\alpha^c\,.
\end{align}
Since the functional derivative also is a tangent vector of the order parameter $n^a$, it can be written as 
\begin{align}\label{lin_com_functional}
    \frac{\delta H}{\delta n^a} = D^\alpha e_\alpha^a
\end{align}
for some coefficients $D^\alpha$ and, by substituting Eq.~\eqref{id_tangent} into Eq.~\eqref{lin_com_functional}, yields Eq.~\eqref{id_functional_derivative}.

The change of the total energy governed by the gLLG is given by
\begin{align}
    \frac{dH}{dt} &= \int_\mathbf{r} \frac{\delta H}{\delta n^a }  \partial_t n^a  \nonumber 
    \\
    &= \int_\mathbf{r} \left[2 f_{abc} \frac{\delta H}{\delta n^a} \frac{\delta H}{\delta n^b} n^c  - \alpha \frac{\delta H}{\delta n^a} \frac{\delta H}{\delta n^a} \right] \nonumber
    \\
    &= - \alpha \int_\mathbf{r} \left| \frac{\delta H}{\delta \mathbf{n}} \right|^2\,.
\end{align}
Manifestly, the total energy decreases in the presence of the Gilbert damping.

\begin{widetext}

\section{Topological current conservation of fractonic nature}
\label{app:topological-current-cons}

Here we show $\partial_\mu J_\mu = 0$ for the topological three-current
{\begin{align}
    J^\mu = \frac{\epsilon^{\mu\nu\lambda}}{8\pi}f_{abc} n^a \partial_\nu n^b \partial_\lambda n^c \,.
\end{align}}
From the definition $n^a = \mathbf{z}^\dagger \lambda_a \mathbf{z}$ with $\mathbf{z} = U \mathbf{z}_0$, derivatives of the field $n^a$ can be expressed as
\begin{align}
\partial_\mu n^a &= {\bf z}^\dag  [ U \partial_\mu U^\dagger, \lambda_a ] {\bf z} 
    = \frac{i}{2} \omega_\mu^b {\bf z}^\dag  [ \lambda_b , \lambda_a  ] {\bf z} = f_{abc}\omega_\mu^b {\bf z}^\dag \lambda_c {\bf z}  = f_{abc} \omega_\mu^b n^c\, ,  \label{variation-n}
\end{align}
by using $U \partial_\mu U^\dag = i \omega_\mu^b \lambda_b /2$ for some $\omega_\mu^b$~\cite{NairQFT}.
The divergence of $J_\mu$ is written as
{\begin{align}
    \partial_\mu J^\mu = \frac{\epsilon^{\mu\nu\lambda}}{8\pi}f_{abc} \partial_\mu n^a \partial_\nu n^b \partial_\lambda n^c \,, 
 \end{align}}
and, using Eq.~\eqref{variation-n} and the contraction rule of the structure constants~\footnote{The identity {$f_{abc}f_{ade} = \frac{2}{N}\left(\delta_{bd}\delta_{ce} - \delta_{be}\delta_{cd}\right) + d_{bda}d_{cea} - d_{bea}d_{cda}$} is equivalent to Eq.~(A6) in Ref.~\cite{Kaplan1967}}, we obtain
{\begin{align}
    \partial_\mu J^\mu &= \frac{\epsilon^{\mu\nu\lambda}}{8\pi}f_{abc} f_{ade} \omega_\mu^d n^e \partial_\nu n^b \partial_\lambda n^c  = {\frac{\epsilon^{\mu\nu\lambda}}{8\pi}\left[\frac{2}{N}\left(\delta_{bd}\delta_{ce} - \delta_{be}\delta_{cd}\right) + d_{bda}d_{cea} - d_{bea}d_{cda}\right] \omega_\mu^d n^e \partial_\nu n^b \partial_\lambda n^c} \nonumber
    \\
    &={\frac{\epsilon^{\mu\nu\lambda}}{8\pi}\left[ d_{bda}d_{cea} - d_{bea}d_{cda}\right] \omega_\mu^d n^e \partial_\nu n^b \partial_\lambda n^c =\frac{\epsilon^{\mu\nu\lambda}}{8\pi}\frac{N-2}{N}\left[ d_{bda} \partial_\nu n^b\partial_\lambda n^a - d_{cda} \partial_\nu n^a  \partial_\lambda n^c\right] \omega_\mu^d} \nonumber
    \\
    &= 0\,,
\end{align}}

Invoking gLL, the continuity equation $\partial_t J_t + \bm\nabla \cdot \mathbf{J} = 0$ can be further written as a fractonic continuity equation:
\begin{align}
    \partial_t J_t &=\frac{\epsilon_{ij}}{4 \pi} f_{abc}\partial_i \left[ n^a \partial_t n^b \partial_j n^c\right] = -\frac{\epsilon_{ij}}{4 \pi} f_{abc} \partial_i \left[  \partial_t n^a n^b \partial_j n^c\right] = -\frac{\epsilon_{ij}}{2 \pi} f_{abc}f_{ade} \partial_i \left[  \frac{\delta H}{\delta n^d} n^e n^b \partial_j n^c\right] \nonumber
    \\
    &=
    {-\frac{\epsilon_{ij}}{2 \pi} \left[\frac{2}{N}\left(\delta_{bd}\delta_{ce} - \delta_{be}\delta_{cd}\right) + d_{bda}d_{cea} - d_{bea}d_{cda}\right] \partial_i \left[  \frac{\delta H}{\delta n^d} n^e n^b \partial_j n^c\right]} \nonumber
    \\
    &={-\frac{\epsilon_{ij}}{2 \pi} \left[-\frac{2}{N}\partial_i \left[  \frac{\delta H}{\delta n^c} n^b n^b \partial_j n^c\right] + (d_{bda}d_{cea} - d_{bea}d_{cda}) \partial_i \left[  \frac{\delta H}{\delta n^d} n^e n^b \partial_j n^c\right]\right]} \nonumber
    \\
    &={\frac{\epsilon_{ij}}{2 \pi} \left[\frac{2}{N}\frac{2(N-1)}{N}\partial_i \left[  \frac{\delta H}{\delta n^c}  \partial_j n^c\right]  + \frac{2(N-2)}{N}d_{cda} \partial_i \left[  \frac{\delta H}{\delta n^d} n^a \partial_j n^c\right] - \frac{(N-2)}{N}d_{bda}\partial_i \left[  \frac{\delta H}{\delta n^d}  n^b \partial_j n^a\right]\right]} \nonumber
    \\
    &={\frac{\epsilon_{ij}}{2 \pi} \left[\frac{2}{N}\frac{2(N-1)}{N}\partial_i \left[  \frac{\delta H}{\delta n^c}  \partial_j n^c\right] +\left(\frac{(N-2)}{N}\right)^2\partial_i \left[  \frac{\delta H}{\delta n^d}   \partial_j n^d\right] \right]} \nonumber
    \\
    &={\frac{\epsilon_{ij}}{2 \pi}\partial_i \left[  \frac{\delta H}{\delta n^d}   \partial_j n^d\right]} \nonumber
    \\
    &={\frac{\epsilon_{ij}}{2 \pi}\partial_i \left[  \left[\frac{\partial \mathcal{H}}{\partial n^d} - \partial_k \left(\frac{\partial \mathcal{H}}{\partial(\partial_k n^d)}\right)\right]\partial_j n^d\right] =\frac{\epsilon_{ij}}{2 \pi}\partial_i \left[  \frac{\partial \mathcal{H}}{\partial n^d} \partial_j n^d +\frac{\partial \mathcal{H}}{\partial(\partial_k n^d)}\partial_j\partial_k n^d - \partial_k \left(\frac{\partial \mathcal{H}}{\partial(\partial_k n^d)}\partial_j  n^d\right)  \right]} \nonumber
    \\
    &={\frac{\epsilon_{ij}}{2 \pi}\partial_i \left[\partial_j\mathcal{H}   - \partial_k \left(\frac{\partial \mathcal{H}}{\partial(\partial_k n^d)}\partial_j  n^d\right) \right] =\frac{\epsilon_{ij}}{2 \pi}\partial_i \left[\partial_k\mathcal{H} \delta_{jk}   - \partial_k \left(\frac{\partial \mathcal{H}}{\partial(\partial_k n^d)}\partial_j  n^d\right) \right]} \nonumber
    \\
    &={\frac{\epsilon_{ij}}{2\pi}\partial_i \partial_k\left[\mathcal{H}\delta_{jk} - \frac{\partial\mathcal{H}}{\partial(\partial_k n^d)} \partial_j n^d\right]} \nonumber
    \\
    &={\frac{\epsilon_{ij}}{2\pi}\partial_i \partial_k\sigma_{jk}}\, . 
\end{align}

In the presence of dissipation, the field $n^a$ obeys the dynamics under the gLLG equation~\eqref{gLLG} and the continuity equation for the \cpn fluid is modified as
\begin{align}
    \partial_t J_t & = -\frac{\epsilon_{ij}}{4 \pi} f_{abc} \partial_i \left[  \partial_t n^a n^b \partial_j n^c\right] = {\frac{\epsilon_{ij}}{2\pi}\partial_i \partial_k\sigma_{jk}} + \alpha \frac{\epsilon_{ij}}{4 \pi} f_{abc} \partial_i \left[ \frac{\delta H}{\delta n^a} n^b \partial_j n^c\right] \,.
\end{align}

\section{Projective-field derivation of the fractonic continuity equation}
\label{app:z-fractonic}
{

This appendix gives the derivation of the fractonic continuity equation directly in the projective-field representation.
The Berry curvature of the $\mathrm{CP}^{N-1}$ representative is
\begin{align}
    F_{\mu\nu}
    &=\partial_\mu a_\nu-\partial_\nu a_\mu 
      =-i\left[(\partial_\mu\mathbf z)^\dagger\partial_\nu\mathbf z
      -(\partial_\nu\mathbf z)^\dagger\partial_\mu\mathbf z\right] \nonumber \\
    &=-i\,\mathrm{Tr}\left(l[\partial_\mu l,\partial_\nu l]\right).
    \label{F-projector-app}
\end{align}
Using $l=\mathbf{1}_{N\times N}/N+n^a\lambda_a/2$ and $[\lambda_b,\lambda_c]=2if_{bcd}\lambda_d$, one obtains
\begin{align}
    \mathrm{Tr}\left(l[\partial_\mu l,\partial_\nu l]\right)
    =\frac{i}{2}f_{abc}n^a\partial_\mu n^b\partial_\nu n^c,
\end{align}
and therefore
\begin{align}
    F_{\mu\nu}=\frac12 f_{abc}n^a\partial_\mu n^b\partial_\nu n^c .
\end{align}
The topological current is consequently
\begin{align}
    J^\mu=\frac{1}{4\pi}\epsilon^{\mu\nu\lambda}F_{\nu\lambda}
    =\frac{1}{8\pi}\epsilon^{\mu\nu\lambda}f_{abc}n^a\partial_\nu n^b\partial_\lambda n^c,
\end{align}
which establishes the equivalence between the $\mathbf z$ and $\mathbf n$ expressions.

We now impose the nondissipative Hamiltonian flow
\begin{align}
    iD_t\mathbf z=P_{\mathbf z}\frac{\delta H}{\delta\mathbf z^\dagger}
\end{align}
for a local, gauge-invariant, spatially translationally invariant Hamiltonian
\begin{align}
    H=\int d^2\mathbf r\,
    \mathcal H\left(\mathbf z,\mathbf z^\dagger,
    D_i\mathbf z,(D_i\mathbf z)^\dagger\right).
\end{align}
Equation~\eqref{F-projector-app} gives
\begin{align}
    F_{tk}
    &=-i\left[(D_t\mathbf z)^\dagger D_k\mathbf z
    -(D_k\mathbf z)^\dagger D_t\mathbf z\right] \\
    &=2\,\mathrm{Re}\left[
    \left(\frac{\delta H}{\delta\mathbf z^\dagger}\right)^\dagger
    D_k\mathbf z\right].
    \label{Ftk-H-app}
\end{align}
Consider the infinitesimal local translation
\begin{align}
    \delta\mathbf z=-\xi^k(\mathbf r)\partial_k\mathbf z,
    \qquad
    \delta\mathbf z^\dagger=-\xi^k(\mathbf r)(\partial_k\mathbf z)^\dagger .
\end{align}
The variational definition of the functional derivative and Eq.~\eqref{Ftk-H-app} give
\begin{align}
    \delta H
    =-2\int d^2\mathbf r\,\xi^k\,
    \mathrm{Re}\left[
    \left(\frac{\delta H}{\delta\mathbf z^\dagger}\right)^\dagger
    D_k\mathbf z\right]
    =-\int d^2\mathbf r\,\xi^k F_{tk}.
    \label{deltaH-Ftk-app}
\end{align}
On the other hand, evaluating the same local translation directly from the Hamiltonian density yields
\begin{align}
    \delta H
    &=\int d^2\mathbf r\,
    \left[
        \mathcal H\delta_{kj}
        -2\,\mathrm{Re}\left(
        \left(\frac{\partial\mathcal H}{\partial(D_j\mathbf z)^\dagger}\right)^\dagger
        D_k\mathbf z
        \right)
    \right]\partial_j\xi^k .
\end{align}
After integrating by parts and comparing with Eq.~\eqref{deltaH-Ftk-app}, the arbitrariness of $\xi^k$ implies
\begin{align}
    F_{tk}=\partial_j\tau_{kj},
    \qquad
    \tau_{kj}=\mathcal H\delta_{kj}
        -2\,\mathrm{Re}\left(
        \left(\frac{\partial\mathcal H}{\partial(D_j\mathbf z)^\dagger}\right)^\dagger
        D_k\mathbf z
        \right).
    \label{tau-z-app}
\end{align}
The spatial topological current can therefore be written as a divergence,
\begin{align}
    J^i=-\frac{1}{2\pi}\epsilon_{ik}F_{tk}
    =\partial_jJ^{ij},
    \qquad
    J^{ij}=-\frac{1}{2\pi}\epsilon_{ik}\tau_{kj}.
\end{align}
Substitution into the ordinary topological continuity equation $\partial_tJ^t+\partial_iJ^i=0$ gives
\begin{align}
    \partial_tJ^t+\partial_i\partial_jJ^{ij}=0,
\end{align}
which is the fractonic continuity equation in the projective-field formulation.
}

\section{GMP algebra}
\label{app:GMP}

In this section, we derive the Poisson bracket of the topological charge densities $\{J_t(\mathbf{r}) , J_t( \mathbf{r}')\}$. From the Poisson bracket of the field $n^a$,
\begin{align}\label{pb}
    \{n^a ( \mathbf{r} ), n^b (\mathbf{r}') \} = 2f_{abc} n^c(\mathbf{r}) \delta(\mathbf{r} - \mathbf{r}')\,,
\end{align}which corresponds to the commutation relation of the $\mathrm{SU}(N)$ generators, the Poisson bracket of the field and its derivative is directly obtained as 
\begin{align}
    \{ \partial_i n^a (\mathbf{r}), n^b(\mathbf{r}') \} &= 2f_{abc} n^c(\mathbf{r}') \partial_i \delta(\mathbf{r}- \mathbf{r}')\,.
\end{align}
The Poisson bracket of the topological charge density and the field is computed by considering the following integral
\begin{align}\label{pb_topn_int}
    \int_{\mathbf{r},\mathbf{r}'} \{ J_t(\mathbf{r}) , n^d (\mathbf{r}') \} g(\mathbf{r}, \mathbf{r}') &= \int_{\mathbf{r},\mathbf{r}'} \frac{\epsilon_{ij}}{8\pi} f_{abc} \{ n^a  \partial_i n^b  \partial_j n^c  , n^d (\mathbf{r}')\} g = \int_{\mathbf{r},\mathbf{r}'} \frac{\epsilon_{ij}}{8\pi} f_{abc} \bigg[\{ n^a    , n^d (\mathbf{r}')\} \partial_i n^b  \partial_j n^c + 2n^a  \partial_i n^b \{   \partial_j n^c  , n^d (\mathbf{r}')\}\bigg] g \,, \nonumber
    \\
    &= \int_{\mathbf{r},\mathbf{r}'} \frac{\epsilon_{ij}}{8\pi} f_{abc} \bigg[3\{ n^a    , n^d (\mathbf{r}')\} \partial_i n^b  \partial_j n^c g - 2n^a  \partial_i n^b \{ n^c  , n^d (\mathbf{r}')\} \partial_j g \bigg] \,,
\end{align}
where $g$ is an arbitrary function depending on $\mathbf{r}$ and $\mathbf{r}'$. Here, for the field variables, only the dependence on $\mathbf{r}'$ is explicitly written, and the argument $\mathbf{r}$ is omitted for brevity. One can show that the first term in the last expression of Eq.~\eqref{pb_topn_int} vanishes from the identities~\eqref{id}. The second term becomes
\begin{align}
    \int_{\mathbf{r},\mathbf{r}'} \frac{\epsilon_{ij}}{4\pi} f_{abc} n^c \partial_i n^b \{ n^a, n^d (\mathbf{r}') \} \partial_j g &= \int_{\mathbf{r},\mathbf{r}'} \frac{\epsilon_{ij}}{2\pi} f_{abc}f_{ade} n^c \partial_i n^b n^e  \delta(\mathbf{r} -\mathbf{r}') \partial_j g \nonumber 
    \\
    &= {\int_{\mathbf{r},\mathbf{r}'} \frac{\epsilon_{ij}}{2\pi} \left[\frac{2}{N}\left(\delta_{bd}\delta_{ce} - \delta_{be}\delta_{cd}\right) + d_{bda}d_{cea} - d_{bea}d_{cda} \right] n^c \partial_i n^b n^e  \delta (\mathbf{r}- \mathbf{r}') \partial_j g} \nonumber
    \\
    &= {\int_{\mathbf{r},\mathbf{r}'} \frac{\epsilon_{ij}}{2\pi} \left[ + \frac{2}{N} \frac{2(N-1)}{N} \partial_i n^d - \frac{N-2}{N} d_{cda} \partial_i n^a n^c + \frac{2(N-2)}{N} d_{bda} \partial_i n^b n^a \right]  \delta (\mathbf{r}-\mathbf{r}') \partial_j g} \nonumber
    \\
    &= {\int_{\mathbf{r},\mathbf{r}'} \frac{\epsilon_{ij}}{2\pi} \partial_i n^d  \delta(\mathbf{r}-\mathbf{r}') \partial_j g} \nonumber
    \\
    &= {-\int_{\mathbf{r},\mathbf{r}'} \frac{\epsilon_{ij}}{2\pi} \partial_i n^d \partial_j \delta(\mathbf{r}-\mathbf{r}') g}\,.
\end{align}
Thus, we have
\begin{align}
    \{J_t(\mathbf{r}),n^d(\mathbf{r}')\} = {-\frac{\epsilon_{ij}}{2\pi} \partial_i n^d(\mathbf{r}) \partial_j \delta(\mathbf{r}- \mathbf{r}')}\,.
\end{align}
For the Poisson bracket of the topological charge densities, we similarly compute the following quantity
\begin{align}\label{poisson_jj}
    \int_{\mathbf{r},\mathbf{r}'} \{ J_t(\mathbf{r}), J_t (\mathbf{r}') \} g(\mathbf{r},\mathbf{r}') &= \int_{\mathbf{r},\mathbf{r}'} \left[ \{n^d, J_t (\mathbf{r}') \} \frac{\partial J_t}{\partial n^d}  + \{ \partial_k n^d , J_t (\mathbf{r}') \} \frac{\partial J_t}{\partial \partial_k n^d}    \right] g \nonumber
    \\
    &= \int_{\mathbf{r},\mathbf{r}'} \{n^d, J_t (\mathbf{r}') \}\left[  \frac{\partial J_t}{\partial n^d}  -  \partial_k\frac{\partial J_t}{\partial \partial_k n^d}   \right] g - \int_{\mathbf{r},\mathbf{r}'}\{n^d, J_t (\mathbf{r}') \} \frac{\partial J_t}{\partial \partial_k n^d} \partial_k g
\end{align}
The first term in the last line of Eq.~\eqref{poisson_jj} gives
\begin{align}\label{jjfirst}
    \int_{\mathbf{r},\mathbf{r}'} \{n^d, J_t (\mathbf{r}') \}\left[  \frac{\partial J_t}{\partial n^d}  -  \partial_k\frac{\partial J_t}{\partial \partial_k n^d}   \right] g & = \int_{\mathbf{r}, \mathbf{r}'} \frac{\epsilon_{ij}}{2\pi} \partial_i' n^d (\mathbf{r}') \partial_j' \delta(\mathbf{r}-\mathbf{r}') \left[  \frac{\partial J_t}{\partial n^d}  -  \partial_k\frac{\partial J_t}{\partial \partial_k n^d}   \right] g \nonumber
    \\
    & =  {-}\int_{\mathbf{r}, \mathbf{r}'} \frac{\epsilon_{ij}}{2\pi} \partial_i' n^d (\mathbf{r}')  \delta(\mathbf{r}-\mathbf{r}') \left[  \frac{\partial J_t}{\partial n^d}  -  \partial_k\frac{\partial J_t}{\partial \partial_k n^d}   \right] \partial_j' g \nonumber
    \\
    & =  {-}\int_{\mathbf{r}, \mathbf{r}'} \frac{\epsilon_{ij}}{2\pi} \partial_i n^d (\mathbf{r})  \delta(\mathbf{r}-\mathbf{r}') \left[  \frac{\partial J_t}{\partial n^d}  -  \partial_k\frac{\partial J_t}{\partial \partial_k n^d}   \right] \partial_j' g \nonumber
    \\
    &={-}\int_{\mathbf{r}, \mathbf{r}'} \frac{\epsilon_{ij}}{2\pi}    \left[  \partial_i n^d \frac{\partial J_t}{\partial n^d}  + \partial_i \partial_k n^d \frac{\partial J_t}{\partial \partial_k n^d}   \right] \delta(\mathbf{r}-\mathbf{r}')\partial_j' g {-} \int_{\mathbf{r},\mathbf{r}'} \frac{\epsilon_{ij}}{2\pi}\partial_i n^d \frac{\partial J_t}{\partial \partial_k n^d} \partial_k\left[ \delta(\mathbf{r}-\mathbf{r}') \partial_j' g \right] \nonumber
    \\
    &={-}\int_{\mathbf{r}, \mathbf{r}'} \frac{\epsilon_{ij}}{2\pi}    \partial_i J_t \delta(\mathbf{r}-\mathbf{r}')\partial_j' g {-} \int_{\mathbf{r},\mathbf{r}'} \frac{\epsilon_{ij}}{2\pi}\partial_i n^d \frac{\partial J_t}{\partial \partial_k n^d} \partial_k\left[ \delta(\mathbf{r}-\mathbf{r}') \partial_j' g \right] \,.
\end{align}
Meanwhile, the second term of Eq.~\eqref{poisson_jj} is
\begin{align}\label{jjsecond}
    - \int_{\mathbf{r},\mathbf{r}'}\{n^d, J_t (\mathbf{r}') \} \frac{\partial J_t}{\partial \partial_k n^d} \partial_k g &= {-}\int_{\mathbf{r},\mathbf{r}'} \frac{\epsilon_{ij}}{2\pi} \partial_i' n^d(\mathbf{r}') \partial_j' \delta (\mathbf{r}-\mathbf{r}') \frac{\partial J_t}{\partial \partial_k n^d } \partial_k g = \int_{\mathbf{r},\mathbf{r}'} \frac{\epsilon_{ij}}{2\pi} \partial_i n^d \delta (\mathbf{r}-\mathbf{r}') \frac{\partial J_t}{\partial \partial_k n^d } \partial_j' \partial_k g\,.
\end{align}
Taking the sum of Eqs.~\eqref{jjfirst} and \eqref{jjsecond}, we can express Eq.~\eqref{poisson_jj} as
\begin{align}
    \int_{\mathbf{r},\mathbf{r}'} \{ J_t (\mathbf{r}), J_t (\mathbf{r}')\} g(\mathbf{r},\mathbf{r}') &={-}\int_{\mathbf{r}, \mathbf{r}'} \frac{\epsilon_{ij}}{2\pi}    \partial_i J_t \delta(\mathbf{r}-\mathbf{r}')\partial_j' g {-} \int_{\mathbf{r},\mathbf{r}'} \frac{\epsilon_{ij}}{2\pi}\partial_i n^d \frac{\partial J_t}{\partial \partial_k n^d} \partial_k\delta(\mathbf{r}-\mathbf{r}') \partial_j' g \nonumber
    \\
    &={-}\int_{\mathbf{r}, \mathbf{r}'} \frac{\epsilon_{ij}}{2\pi}    \partial_i J_t \partial_j \delta(\mathbf{r}-\mathbf{r}') g {-} \int_{\mathbf{r},\mathbf{r}'} \frac{\epsilon_{ij}}{2\pi}\partial_i n^d \frac{\partial J_t}{\partial \partial_k n^d} \partial_j \partial_k\delta(\mathbf{r}-\mathbf{r}') g\nonumber 
    \\
    &={-}\int_{\mathbf{r}, \mathbf{r}'} \frac{\epsilon_{ij}}{2\pi}    \partial_i J_t \partial_j \delta(\mathbf{r}-\mathbf{r}') g {-} \int_{\mathbf{r},\mathbf{r}'} \frac{\epsilon_{ij}}{2\pi}\frac{\epsilon_{kl}}{4\pi} f_{adc} n^a \partial_l n^c \partial_i n^d \partial_j \partial_k\delta(\mathbf{r}-\mathbf{r}') g\nonumber 
    \\
    &={-}\int_{\mathbf{r}, \mathbf{r}'} \frac{\epsilon_{ij}}{2\pi}    \partial_i J_t \partial_j \delta(\mathbf{r}-\mathbf{r}') g {-} \int_{\mathbf{r},\mathbf{r}'} \frac{f_{adc}}{8\pi^2}\left[ n^a \partial_x n^c \partial_y n^d \partial_x \partial_y \delta (\mathbf{r}-\mathbf{r}') + n^a \partial_y n^c \partial_x n^d \partial_y \partial_x \delta (\mathbf{r}-\mathbf{r}') \right]  g\nonumber 
    \\
    &={-}\int_{\mathbf{r}, \mathbf{r}'} \frac{\epsilon_{ij}}{2\pi}    \partial_i J_t \partial_j \delta(\mathbf{r}-\mathbf{r}') g\,.
\end{align}
Therefore, the Poisson bracket of the topological charge densities is given as the long-wavelength limit of the GMP algebra.
\begin{align}\label{pb_toptop}
    \{J_t(\mathbf{r}),J_t(\mathbf{r}')\} = {-\frac{\epsilon_{ij}}{2\pi} \partial_i J_t(\mathbf{r}) \partial_j \delta(\mathbf{r}- \mathbf{r}')}\,.
\end{align}

\section{CP$^2$ skyrmion}
\label{app:CP2-algebra}
The spin-1 state is spanned by three angular momentum states $|1\rangle, |0\rangle, |-1\rangle$, or by the Cartesian basis $\{ |x\rangle, |y\rangle, |z\rangle \}$~\cite{akagi21, shannon22}:
\begin{align} |x\rangle  = i (|1\rangle - |-1\rangle)/\sqrt{2}, \quad
|y\rangle  = (|1\rangle + |-1\rangle)/\sqrt{2}, \quad
|z\rangle  = -i |0\rangle .\label{cartesian basis} \end{align} 
The eight generators of SU(3), $\lambda_a$ ($a=1,\cdots,8$) are related to the three magnetic and five quadrupole moments of spin-1 through
\begin{align}\label{quadrupole_def}
\begin{pmatrix}
    \lambda^7 \\
    \lambda^5 \\
    \lambda^2
\end{pmatrix}
=
\begin{pmatrix}
    S^x\\
    -S^y\\
    S^z
\end{pmatrix}
,\quad
-\begin{pmatrix} 
    \lambda^3\\
    \lambda^8\\
    \lambda^1\\
    \lambda^4\\
    \lambda^6\\
\end{pmatrix}
=
\begin{pmatrix}
    Q^{x^2 -y^2}\\
    Q^{3z^2-{\bf r}^2}\\
    Q^{xy}\\
    Q^{xz}\\
    Q^{yz}
\end{pmatrix}=
 \begin{pmatrix}
    (S^x)^2-(S^y)^2\\
    \frac{1}{\sqrt{3}}[\mathbf{S}\cdot\mathbf{S} - 3(S^z)^2]\\
    S^xS^y + S^yS^x\\
    S^zS^x + S^xS^z\\
    S^yS^z + S^zS^y 
\end{pmatrix} , 
\end{align}
For a given $\bf z$, the magnetic dipole and quadrupole components can be worked out as $n^a = {\bf z}^\dag \lambda_a {\bf z} = z^*_i (\lambda_a )_{ij} z_j$, with the components $z_j$ given in the Cartesian basis $|{\bf z} \rangle = \sum_{j=x,y,z} z_j |j \rangle$.

For $\bf z$ given in the exact solution of the $\mathrm{CP}^2$ skyrmion [Eq.~\eqref{general_solution}], the magnetic dipole and quadrupole moments are
\begin{align}\label{spin-and-quadrupole-average}
    \langle \mathbf{S} \rangle &= \frac{i}{\Delta}\left(\xi^{2|n|}({\bf u}\times{\bf u}^*)
    +\xi^{|n|}r^{|n|}e^{in\theta}({\bf v} \times {\bf u}^*)
    +\xi^{|n|}r^{|n|}e^{-in\theta}({\bf u} \times {\bf v}^*)
    +r^{2|n|}({\bf v} \times {\bf v}^*)\right)\,, \nonumber
    \\
    \langle Q^{x^2-y^2} \rangle &= - \frac{1}{\Delta}\left(\xi^{2|n|}(u_1^{*}u_1-u_2^{*}u_2) + \xi^{|n|}r^{|n|}e^{in\theta}(u_1^{*}v_1-u_2^{*}v_2) + \xi^{|n|}r^{|n|}e^{-in\theta}(v_1^{*}u_1-v_2^{*}u_2) + r^{2|n|}(v_1^{*}v_1-v_2^{*}v_2) \right)\,,\nonumber
    \\
    \langle Q^{3z^2-{\bf r}^2} \rangle &= -\frac{1}{\Delta\sqrt{3}}\left( \xi^{2|n|}(|{\bf u}|^2 - 3u_3^{*}u_3) + \xi^{|n|}r^{|n|}e^{in\theta}({\bf u}^* \cdot{\bf v}- 3u_3^{*}v_3) + \xi^{|n|}r^{|n|}e^{-in\theta}({\bf v}^* \cdot {\bf u} - 3v_3^{*}u_3) + r^{2|n|}(|{\bf v}|^2 - 3v_3^{*}v_3)  \right)\,,\nonumber
    \\
    \langle Q^{xy} \rangle &= -\frac{1}{\Delta}\left(\xi^{2|n|}(u_2^{*}u_1+u_1^{*}u_2) + \xi^{|n|}r^{|n|}e^{in\theta}(u_2^{*}v_1+u_1^{*}v_2) + \xi^{|n|}r^{|n|}e^{-in\theta}(v_2^{*}u_1 + v_1^{*}u_2) + r^{2|n|}(v_2^{*}v_1 + v_1^{*}v_2) \right)\,, \nonumber
    \\
    \langle Q^{xz} \rangle &= -\frac{1}{\Delta}\left(\xi^{2|n|}(u_3^{*}u_1+u_1^{*}u_3) + \xi^{|n|}r^{|n|}e^{in\theta}(u_3^{*}v_1+u_1^{*}v_3) + \xi^{|n|}r^{|n|}e^{-in\theta}(v_3^{*}u_1 + v_1^{*}u_3) + r^{2|n|}(v_3^{*}v_1 + v_1^{*}v_3) \right)\,,\nonumber
    \\
    \langle Q^{yz} \rangle &= -\frac{1}{\Delta}\left(\xi^{2|n|}(u_3^{*}u_2+u_2^{*}u_3) + \xi^{|n|}r^{|n|}e^{in\theta}(u_3^{*}v_2+u_2^{*}v_3) + \xi^{|n|}r^{|n|}e^{-in\theta}(v_3^{*}u_2 + v_2^{*}u_3) + r^{2|n|}(v_3^{*}v_2 + v_2^{*}v_3) \right)\,,
\end{align}
where $\Delta = \xi^{2|n|}+r^{2|n|}$. The scalar spin chirality is given by
\begin{align}
\langle {\bf S} \rangle \cdot \partial_x \langle{\bf S}\rangle \times \partial_y \langle {\bf S}\rangle
=& \frac{2n|n| (r\xi)^{2|n|}}{r^2 \Delta^3}
\bigg[r^{2|n|}|\mathbf{u}\cdot(\mathbf{v}\times\mathbf{v}^*)|^2+\xi^{2|n|}|\mathbf{v}\cdot(\mathbf{u}\times\mathbf{u}^*)|^2 \nonumber
    \\
    &-(r\xi)^{|n|}e^{-in\theta}({\bf v} \times {\bf v}^*) \cdot \left(({\bf u} \times {\bf v}^*) \times ({\bf u} \times {\bf u}^*)\right)+(r\xi)^{|n|}e^{in\theta}({\bf v} \times {\bf v}^*) \cdot \left(({\bf v} \times {\bf u}^*) \times ({\bf u} \times {\bf u}^*)\right)
    \bigg]\label{cp1 density}.
\end{align}
The norm squared of the magnetic dipole moment is worked out to be
\begin{align}\label{normsquared}
    \langle \mathbf{S} \rangle \cdot \langle\mathbf{S} \rangle =-\frac{1}{\Delta^2} \bigg[ & \xi^{4|n|} (\mathbf{u}\times \mathbf{u}^*)\cdot (\mathbf{u}\times \mathbf{u}^*) + r^{4|n|} (\mathbf{v} \times \mathbf{v}^*) \cdot (\mathbf{v} \times \mathbf{v}^*) + 2\xi^{2|n|} r^{2|n|} \left(|\mathbf{u}\cdot \mathbf{v}|^2-1\right) \nonumber
    \\
    & + 2\xi^{3|n|} r^{|n|} \left( e^{in\theta} (\mathbf{u}\cdot \mathbf{v}) (\mathbf{u}^* \cdot \mathbf{u}^*) + e^{-in\theta} (\mathbf{u}\cdot \mathbf{u}) (\mathbf{u}^* \cdot \mathbf{v}^*)\right)
    + 2\xi^{|n|} r^{3|n|} \left( e^{in\theta} (\mathbf{v}\cdot \mathbf{v}) (\mathbf{u}^* \cdot \mathbf{v}^*) + e^{-in\theta} (\mathbf{u}\cdot \mathbf{v}) (\mathbf{v}^* \cdot \mathbf{v}^*)\right) \nonumber
    \\
    & + 2\xi^{2|n|} r^{2|n|}\left( e^{i2n\theta} (\mathbf{v}\cdot \mathbf{v})  (\mathbf{u}^*\cdot \mathbf{u}^*) + e^{-i2n\theta} (\mathbf{u}\cdot \mathbf{u})(\mathbf{v}^*\cdot \mathbf{v}^*) \right) \bigg]\,.
\end{align}

\section{Energy minimization for the CP$^2$ skyrmion}
\label{app:energy-analysis}
{
We first evaluate the two additional contributions in Eq.~\eqref{energy_with_quadratic} within the BPS ansatz. The internal vectors $({\bf u},{\bf v})$ are subject to the normalization and BPS orthogonality conditions
}
\begin{align}
    {
    {\bf u}^* \cdot{\bf u}={\bf v}^* \cdot{\bf v}=1,
    \qquad
    {\bf u}^* \cdot{\bf v}=0.
    }
\end{align}

By decomposing the complex vectors into their real and imaginary parts, $\mathbf{u}=\mathbf{u}_r+i\mathbf{u}_i$ and $\mathbf{v}=\mathbf{v}_r+i\mathbf{v}_i$, the {anisotropy contribution} to the energy~\eqref{energy_with_quadratic} can be expressed, after integration over the polar angle, as
\begin{align}\label{norm_integral}
    \kappa\int_\mathbf{r}\langle\mathbf{S}\rangle\cdot\langle\mathbf{S}\rangle
    =\kappa\int\frac{2\pi r\,dr}{(\xi^{2|n|}+r^{2|n|})^2}
    \bigg[
    \xi^{4|n|}4|\mathbf{u}_r\times\mathbf{u}_i|^2
    +r^{4|n|}4|\mathbf{v}_r\times\mathbf{v}_i|^2
    -2\xi^{2|n|}r^{2|n|}\left(|\mathbf{u}\cdot\mathbf{v}|^2-1\right)
    \bigg].
\end{align}
For $\kappa>0$, one can show that the vectors minimizing the energy functional are given by
\begin{align}
    \mathbf{u} = e^{i\frac{\alpha}{2}} \hat{\mathbf{e}}\,, \quad \mathbf{v} = e^{i\frac{\beta}{2}}\hat{\mathbf{e}}\,,
\end{align}
with two arbitrary phases $\alpha,\beta$, and an arbitrary real unit vector $\hat{\mathbf{e}}$. { Consequently, ${\bf u}^* \cdot{\bf v} = e^{i(\beta-\alpha)/2}$, whose magnitude is unity. This contradicts the BPS constraint ${\bf u}^* \cdot{\bf v}=0$.
Hence, no purely quadrupolar solution exists within the BPS ansatz for $\kappa>0$.} In the case of $\kappa<0$, there exists no pair of constant orthonormal vectors $(\mathbf{u},\mathbf{v})$
minimizing all terms in the integrand of Eq.~\eqref{norm_integral}. Since the characteristic length $\xi$ is much smaller than the system size, the contribution of the term proportional to $\xi^{4|n|}$ to the total energy can be neglected. From the second and third terms of Eq.~\eqref{norm_integral}, we find the following relations
\begin{align}
    |\mathbf{v}_r| = |\mathbf{v}_i| = \frac{1}{\sqrt{2}}\,, \quad \mathbf{v}_r \cdot \mathbf{v}_i = 0 \,, \quad \mathbf{u} = 2 e^{i \varphi} \mathbf{v}_r \times \mathbf{v}_i 
\end{align}
where $\varphi$ is an arbitrary phase. The quartet of real vectors $\{\mathbf{v}_r, \mathbf{v}_i, \mathbf{u}_r, \mathbf{u}_i\}$ forms a rigid body which is parameterized by $\mathrm{SO}(3)$ rotation, and the compleax vector $\mathbf{u}$ has $\mathrm{U}(1)$ degeneracy for given $\mathbf{v}_r$ and $\mathbf{v}_i$. Thus the configuration is parameterized by the $\mathrm{SO}(3)$ rotation $\mathcal{R}(\alpha,\beta,\gamma) = e^{\alpha L_z}e^{\beta L_y}e^{\gamma L_z}$, where $L_i$ are the generators of $\mathrm{SO}(3)$, and the $\mathrm{U}(1)$ rotation $e^{i\varphi}$ as
\begin{align}\label{so3u1}
    {\bf v}_r &= \mathcal{R}(\alpha, 
    \beta, \gamma)
    \begin{pmatrix}
        \frac{1}{\sqrt{2}}\\
        0\\
        0
    \end{pmatrix}=\frac{1}{\sqrt{2}}
    \begin{pmatrix}
        \cos{\alpha}\cos{\beta}\cos{\gamma}-\sin{\alpha}\sin{\gamma}\\
        \cos{\beta}\cos{\gamma}\sin{\alpha}+\cos{\alpha}\sin{\gamma}\\
        -\cos{\gamma}\sin{\beta}
    \end{pmatrix}\,, \nonumber
    \\
    {\bf v}_i &= \mathcal{R}(\alpha, 
    \beta, \gamma)
    \begin{pmatrix}
        0\\
        \frac{1}{\sqrt{2}}\\
        0
    \end{pmatrix}=\frac{1}{\sqrt{2}}
    \begin{pmatrix}
        -\cos{\alpha}\cos{\beta}\sin{\gamma}-\cos{\gamma}\sin{\alpha}\\
        -\cos{\beta}\sin{\alpha}\sin{\gamma}+
        \cos{\alpha}\cos{\gamma}\\
        \sin{\beta}\sin{\gamma}
    \end{pmatrix}\,, \nonumber
    \\
    {\bf v} &= \frac{e^{-i\gamma}}{\sqrt{2}}
    \begin{pmatrix}
        \cos{\alpha}\cos{\beta}-i\sin{\alpha}\\
        \sin{\alpha}\cos{\beta}+i\cos{\alpha}\\
        -\sin{\beta}
    \end{pmatrix}\,, \quad 
    {\bf u} = e^{i \varphi}
    \begin{pmatrix}
        \cos{\alpha}\sin{\beta}\\
        \sin{\alpha}\sin{\beta}\\
        \cos{\beta}
    \end{pmatrix}\,.
\end{align}
Employing this solution, the magnetic dipole moments from Eq.~\eqref{spin-and-quadrupole-average} are expressed as
\begin{align}
    \langle S^x \rangle  &= \frac{-\sqrt{2}\xi^n r^n(\sin{\alpha}\sin{\psi}+\cos{\alpha}\cos{\beta}\cos{\psi})+r^{2n}\cos{\alpha}\sin{\beta}}{\xi^{2|n|}+r^{2|n|}}\,, \nonumber
    \\
     \langle S^y \rangle  &=     
    \frac{\sqrt{2}\xi^n r^n (\cos{\alpha}\sin{\psi}-\sin{\alpha}\cos{\beta}\cos{\psi})+r^{2n}\sin{\alpha}\sin{\beta}}{\xi^{2|n|}+r^{2|n|}}\,, \nonumber
    \\
    \langle S^z \rangle  &= 
    \frac{\sqrt{2}\xi^n r^n \sin{\beta}\cos{\psi}+r^{2n}\cos{\beta}}{\xi^{2|n|}+r^{2|n|}}\,,
\end{align}
similarly, the magnetic quadrupole moments can be recast as follows:
\begin{align}
    \langle Q^{x^2-y^2} \rangle &= \frac{\sin\beta}{2(r^{2|n|} + \xi^{2|n|})}\left[ (r^{2|n|} - 2\xi^{2|n|})\cos 2 \alpha \sin \beta - 2\sqrt{2} (r\xi)^{|n|} (\cos 2 \alpha \cos \beta \cos \psi + \sin 2 \alpha \sin \phi )
    \right]\,, \nonumber
    \\
    \langle Q^{3z^2-r^2} \rangle &= \frac{-1}{4\sqrt{3}(r^{2|n|} + \xi^{2|n|})}\left[ (r^{2|n|} - 2 \xi^{2|n|}) (1+3 \cos 2 \beta) + 6 \sqrt{2} (r\xi)^{|n|} \cos \psi \sin 2 \beta
    \right] \,, \nonumber
    \\
    \langle Q^{xy} \rangle &= \frac{\sin \beta}{2(r^{2|n|} + \xi^{2|n|})}\left[ (r^{2|n|} - 2 \xi^{2|n|})\sin 2 \alpha \sin \beta  - 2 \sqrt{2} (r\xi)^{|n|} (\cos \beta \cos \psi \sin 2 \alpha - \cos 2\alpha \sin \psi)
    \right] \,, \nonumber
    \\
    \langle Q^{xz} \rangle &= \frac{1}{2(r^{2|n|} + \xi^{2|n|})}\left[ (r^{2|n|} - 2 \xi^{2|n|})\cos\alpha \sin 2\beta  - 2 \sqrt{2} (r\xi)^{|n|} (\cos\alpha \cos 2\beta \cos\psi + \cos\beta \sin \alpha \sin \psi)
    \right] \,, \nonumber
    \\
    \langle Q^{yz} \rangle &= \frac{1}{2(r^{2|n|} + \xi^{2|n|})}\left[ (r^{2|n|} - 2 \xi^{2|n|})\sin\alpha \sin 2\beta  - 2 \sqrt{2} (r\xi)^{|n|} (\sin\alpha \cos 2\beta \cos\psi - \cos\beta \cos \alpha \sin \psi)
    \right] \,,
\end{align}
where $\psi=n\theta-\varphi-\gamma$.

{

Although the anisotropy term determines the optimal internal vectors $({\bf u},{\bf v})$, it does not by itself determine a finite skyrmion radius. We therefore examine the scale dependence of the total energy within the BPS variational ansatz. As shown above, within this ansatz the quartic derivative contribution is independent of $({\bf u},{\bf v})$, while its dependence on the scale $\xi$ competes with that of the anisotropy energy. The optimal radius can therefore be determined by minimizing the $\xi$-dependent part of the total energy.

For the ferromagnetic CP$^2$ skyrmion,
\begin{align}
    {\bf S}^2(\xi)
    =
    \frac{r^{2|n|}(r^{2|n|}+2\xi^{2|n|})}
    {(r^{2|n|}+\xi^{2|n|})^2}.
\end{align}
The anisotropy contribution can then be written as
\begin{align}
    E_0
    &=\kappa\int_{\bf r}{\bf S}^2(\xi)\nn
    &=\kappa\int_{\bf r}
    \left(
    \frac{r^{2|n|}(r^{2|n|}+2\xi^{2|n|})}
    {(r^{2|n|}+\xi^{2|n|})^2}
    \right)\nn
    &=\kappa\int_{\bf r}
    \left(
    1-\frac{\xi^{4|n|}}
    {(r^{2|n|}+\xi^{2|n|})^2}
    \right).
\end{align}
Since $\kappa<0$, this becomes
\begin{align}
    E_0
    &=
    -|\kappa|\int_{\bf r}1
    +
    |\kappa|\int_{\bf r}
    \frac{\xi^{4|n|}}
    {(r^{2|n|}+\xi^{2|n|})^2}\nn
    &=
    \frac{\pi|\kappa|\xi^2}{|n|}
    \Gamma\left(\frac{1}{|n|}\right)
    \Gamma\left(2-\frac{1}{|n|}\right)
    +\mathrm{const.},
\end{align}
where the constant term is independent of $\xi$ and therefore does not affect the minimization with respect to the skyrmion radius.

The quartic derivative contribution can be obtained as
\begin{align}
    E_4
    =
    \eta\int_{\bf r}
    \left[
    \sum_{i=x,y}(D_i{\bf z})^\dagger D_i{\bf z}
    \right]^2=8\pi\eta|n|^4\xi^{4|n|}\int_0^\infty\frac{r^{4|n|-3}\,dr}{(\xi^{2|n|}+r^{2|n|})^4}
    =
    \frac{2\pi\eta|n|^3}{3\xi^2}
    \Gamma\left(2-\frac{1}{|n|}\right)
    \Gamma\left(2+\frac{1}{|n|}\right).
\end{align}
Since the exchange energy is independent of $\xi$ within the BPS family, the $\xi$-dependent part of the total energy is
\begin{align}\label{xi-dependent-energy}
    \Delta E(\xi)
    &=
    E_0+E_4\nn
    &=
    \frac{\pi|\kappa|\xi^2}{|n|}
    \Gamma\left(\frac{1}{|n|}\right)
    \Gamma\left(2-\frac{1}{|n|}\right)
    +
    \frac{2\pi\eta|n|^3}{3\xi^2}
    \Gamma\left(2-\frac{1}{|n|}\right)
    \Gamma\left(2+\frac{1}{|n|}\right)
    +\mathrm{const.}
\end{align}
Minimizing the energy with respect to $\xi$ gives
\begin{align}
    \frac{d\Delta E}{d\xi}
    &=
    \frac{2\pi|\kappa|\xi}{|n|}
    \Gamma\left(\frac{1}{|n|}\right)
    \Gamma\left(2-\frac{1}{|n|}\right)
    -
    \frac{4\pi\eta|n|^3}{3\xi^3}
    \Gamma\left(2-\frac{1}{|n|}\right)
    \Gamma\left(2+\frac{1}{|n|}\right)
    =0.
\end{align}
The optimal scale parameter $\xi_*$ follows the equation
\begin{align}
    \xi_*^4
    =
    \frac{2\eta|n|^4}{3|\kappa|}
    \frac{
    \Gamma\left(2+\frac{1}{|n|}\right)
    }{
    \Gamma\left(\frac{1}{|n|}\right)
    }\,,
\end{align} and therefore
\begin{align}\label{optimal-radius}
    \xi_\ast
    =
    \left[
    \frac{2\eta|n|^2(|n|+1)}
    {3|\kappa|}
    \right]^{1/4}.
\end{align}
To verify that $\xi_*$ corresponds to a minimum, we calculate the second-order derivative
\begin{align}
    \frac{d^2\Delta E}{d\xi^2}
    &=
    \frac{2\pi|\kappa|}{|n|}
    \Gamma\left(\frac{1}{|n|}\right)
    \Gamma\left(2-\frac{1}{|n|}\right)
    +
    \frac{4\pi\eta|n|^3}{\xi^4}
    \Gamma\left(2-\frac{1}{|n|}\right)
    \Gamma\left(2+\frac{1}{|n|}\right).
\end{align}
Both terms are positive for $\xi>0$, $\eta>0$, and nonzero integer $n$. Therefore,
\begin{align}
    \left.
    \frac{d^2\Delta E}{d\xi^2}
    \right|_{\xi=\xi_\ast}>0,
\end{align}
and $\xi_\ast$ is an energy minimum.
}

\end{widetext}

\bibliography{skyrmion-fracton}

\end{document}